\documentclass[twocolumn,showpacs,10pt,amsmath,amssymb]{revtex4}
\usepackage{graphicx}

\usepackage{amsmath,amsfonts,amssymb,bm}

\begin{document}

\draft
\title{Anomalous diffusion in nonhomogeneous media: Time-subordinated Langevin equation approach}

\author
{Tomasz Srokowski}

\affiliation{
 Institute of Nuclear Physics, Polish Academy of Sciences, PL -- 31-342
Krak\'ow,
Poland }

\date{\today}

\begin{abstract}
Diffusion in nonhomogeneous media is described by a dynamical process driven by a general L\'evy 
noise and subordinated to a random time; 
the subordinator depends on the position. This problem is approximated by a multiplicative process 
subordinated to a random time: it separately takes into account effects related to the medium structure 
and the memory. Density distributions and moments are derived from the solutions of 
the corresponding Langevin equation and compared with the numerical calculations for the exact problem. 
Both subdiffusion and enhanced diffusion are predicted. Distribution 
of the process satisfies the fractional Fokker-Planck equation. 
\end{abstract} 


\maketitle



A typical linear dependence on the variance on time, which follows from 
the standard central limit theorem and leads to normal diffusion, may be no longer valid when 
we deal with a transport on complicated structures. That happens, and deviations 
from the Fick law emerge, in the presence of long-range correlations, wide distributions 
and memory effects \cite{bou}. Impurities and regular structures in the medium cause a bias 
for transport: the particle may be trapped for some time or accelerated. 
The dynamics becomes non-Markovian when one assumes that the waiting-time distribution falls slowly, 
$\sim t^{-\beta-1}$ ($0<\beta<1$); it is governed by the continuous time random walk theory (CTRW) which 
predicts a subdiffusive behaviour \cite{met}. In an equivalent formulation, dynamics is Markovian and 
governed by an ordinary Langevin equation but it proceeds in an auxiliary, operational time instead 
of the physical time. Then this process is subordinated to that in the physical time which is 
a random quantity defined by a broad distribution, usually in a form of the stable, one-sided 
L\'evy distribution with a stability index $\beta$, and this subordinator introduces memory effects \cite{fog}. 
Long tails of the random time distribution determine the anomalous diffusion law: $\langle x^2\rangle\sim t^\beta$. 
Random force in the Langevin equation may be Gaussially distributed but may also obey general 
L\'evy statistics with a stability index $0<\alpha<2$. Such processes exhibit long jumps (L\'evy flights) 
making the variance divergent; they are frequently observed in many areas of science \cite{klag}. 
An interesting example of L\'evy flights has been recently analysed in an experimental study of cracking 
of the heterogeneous materials \cite{tall}. 

The traditional CTRW used to be applied to the transport on nonhomogeneous media but that nonhomogeneity 
is reduced, in fact, to a homogeneous distribution of the noise activation times. One can expect 
that a consistent description of such transport takes into account that the waiting time explicitly 
depends on the position since the trap properties must reflect the medium structure. 
Diffusion on fractals and multifractals provides an example \cite{sch}. 
One possible way to describe diffusion in a composite medium is the introduction of 
position-dependent subdiffusion exponents \cite{che,sti}. On the other hand, 
nonhomogeneous systems are not only characterised by subdiffusion, related to traps, 
but also enhanced diffusion is observed just as a result of the disorder. For example, 
movement of particles between two neighbouring lattice sites in an interacting particle system 
is superdiffusive due to the disorder; it appears subdiffusive in its absence \cite{ben}. 


In this paper, we relate the memory effects connected with anomalous transport 
in a nonhomogeneous medium to the medium structure. This aim is 
accomplished by introducing a function $g(x)$ which serves as a variable intensity of the random time 
distribution $\xi$ and models the position of structures responsible for either trapping or accelerating 
the particle. Then the time-subordinator determined by $\xi$ may be enlarged or decreased, 
according to the medium properties. The problem may be formulated by means of the following set 
of the Langevin equations, 
\begin{eqnarray}
\label{la}
dx(\tau)&=&\eta(d\tau)\nonumber\\
dt(\tau)&=&g(x)\xi(d\tau), 
\end{eqnarray}
where the positive function $g(x)$ is dimensionless and $\eta$ stands for a white noise 
the increments of which are given by a symmetric stable L\'evy 
distribution $L(x;\alpha,0)$. $d\xi(\tau)$, in turn, is a stochastic process given by a one-sided distribution; 
the physical time $t$ is then a rising function of $\tau$. Large values of $g(x)$ correspond 
to the trap positions since then the physical time $t$ rises faster than it would in the case of 
a homogeneous medium and the average trapping time for those positions is longer. 
On the other hand, particle is likely to change its position instantaneously when $g(x)$ is small. 
Eq.(\ref{la}) describes non-Markovian dynamics if $d\xi$ is characterised by long tails. 

Eq.(\ref{la}) appears simple if $\xi$ has a finite first moment $\langle\xi\rangle$; 
then we can approximate $\xi$ by its average, $\xi\longrightarrow \langle\xi\rangle$, 
which yields $\xi(d\tau)=\langle\xi\rangle d\tau$ and 
$\Delta\tau=(\langle\xi\rangle g(x))^{-1}\Delta t$. Taking into account that the first equation (\ref{la}) 
can be discretized as $\Delta x=\Delta \eta\Delta\tau^{1/\alpha}$, we reduce the system (\ref{la}) 
to a single equation with a multiplicative noise in the It\^o interpretation, 
\begin{equation}
\label{mult}
dx(t)=\nu(x)^{1/\alpha}\eta(dt), 
\end{equation}
where $\nu(x)=(\langle\xi\rangle g(x))^{-1}$. Eq.(\ref{mult}) applies, in particular, when the relation 
between both times is deterministic. The probability density distribution 
is given by the fractional Fokker-Planck equation \cite{sch}, 
\begin{equation}
\label{frace}
\frac{\partial p_0(x,t)}{\partial t}=
\frac{\partial^\alpha[\nu(x) p_0(x,t)]}{\partial|x|^\alpha}.
\end{equation} 
Eq.(\ref{frace}) describes Markovian dynamics in the framework of CTRW where the waiting-time 
distribution is Poissonian with a rate $\nu(x)$ \cite{sro06}. Since the rate is position-dependent, 
that approach is able to take into account the medium structure. In the following, we assume 
$\nu(x)=K|x|^{-\theta}$; the constant $K$ was introduced for dimensional 
reasons and will be dropped in the following. The power-law form of $\nu(x)$ is natural for problems 
exhibiting self-similarity which often happens for disordered materials. 
This form of the diffusion coefficient was assumed to describe e.g. diffusion on fractals \cite{osh}, 
turbulent two-particle diffusion and transport of fast electrons in a hot plasma \cite{ved}. 

Eq.(\ref{frace}) can be solved in the diffusion limit 
of small wave numbers for $\theta>-\alpha$; the solution obeys a power-law asymptotics, $\sim|x|^{-1-\alpha}$, 
but its shape at small $|x|$ is not unique. For $\theta<1$, it assumes a form of the stable distribution \cite{sro14}, 
\begin{equation}
\label{sol0}
p_0(x,t)={\cal N}L((At)^{-1/(\alpha+\theta)}|x|;\alpha,0), 
\end{equation}
where $A=\frac{2(\alpha+\theta)}{\pi\alpha^2}\Gamma(\theta/\alpha)\Gamma(1-\theta)
\sin\left(\frac{\pi\theta}{2}\right)$. $p_0(x,t)$ has a characteristic function 
$\widetilde p_0(k,t)=\exp(-ct^{c_\theta}|k|^\alpha)$ where $c_\theta=\alpha/(\alpha+\theta)$ and 
$c=A^{c_\theta}$. Moments of the order $\delta<\alpha$ are given by 
\begin{equation}
\label{mom0}
\langle|x|^\delta\rangle(t)=-\frac{2A^{\delta/(\alpha+\theta)}}{\pi\alpha}\Gamma(-\delta/\alpha)\Gamma(1+\delta)
\sin\left(\frac{\pi\delta}{2}\right)t^{\delta/(\alpha+\theta)}. 
\end{equation}
If $\theta\ge1$, the solution cannot be expressed by the stable distribution but the asymptotics is the same and 
the above time-dependence of the moments is still valid \cite{sro06}. 
For $\alpha=2$ and $\theta>-1$, the solution is given by a stretched-exponential function \cite{osh}, 
\begin{equation}
\label{sol20}
p_0(x,t)={\cal N}t^{-\frac{1+\theta}{2+\theta}} |x|^\theta\exp(-2|x|^{2+\theta}/(2+\theta)^2t), 
\end{equation}
and the variance, 
\begin{equation}
\label{var2}
\langle x^2\rangle(t)\sim t^{2/(2+\theta)}, 
\end{equation}
follows from a direct integration \cite{sro06}. 


Eq.(\ref{la}) comprises two physical effects which are decisive for 
the waiting-time characteristics of the system. They are coupled but 
we will consider them separately and approximate the exact solution of Eq.(\ref{la}) by their 
superposition. First, the nonhomogeneity of the medium structure can be 
taken into account in such a way that the jumping rate in the underlying CTRW depends 
-- deterministically -- on the position and produces a nonlinear time-dependence either of 
the fractional moments or of the variance (for $\alpha=2$). 
The second effect involves memory and originates from long tails of the waiting time distribution; 
it also results in anomalous transport. Superposition consists in a subordination of $x(\tau)$, defined by 
Eq.(\ref{mult}), to a position-independent random time. Therefore, the time lag imposed on the dynamics 
by the position-independent random subordinator is weighted 
by the relative probability of finding the particle at a given position, governed by $g(x)$. 
The above procedure can be formalised as a set of two Langevin equations, 
\begin{eqnarray}
\label{las}
dx(\tau)&=&\nu(x)^{1/\alpha}\eta(d\tau)\nonumber\\
dt(\tau)&=&\xi(d\tau). 
\end{eqnarray}

The statistics of random time is governed by a one-sided, maximally asymmetric stable L\'evy distribution 
$\bar h(\tau,t)=L(\tau;\beta,-\beta)$, where $0<\beta<1$, which vanishes for $\tau<0$. 
A convenient representation of the inverse distribution, 
$h(\tau,t)=\frac{t}{\beta\tau}L(\frac{t}{\tau^{1/\beta}};\beta,-\beta)$ \cite{pir}, 
is the Fox function \cite{non}, 
\begin{eqnarray} 
\label{htt}
h(\tau,t)=\frac{1}{\beta\tau}H_{1,1}^{1,0}\left[\frac{\tau^{1/\beta}}{t}\left|\begin{array}{l}
(1,1)\\
\\
(1,1/\beta)
\end{array}\right.\right]. 
\end{eqnarray}  
Moments of the order $\delta$, both integer and fractional, can be easily evaluated as a Mellin transform 
from Eq.(\ref{htt}) and they are finite for any $\delta>0$. The density distribution 
corresponding to Eq.(\ref{las}) is a joint density 
\begin{equation}
\label{inte}
p(x,t)=\int_0^\infty p_0(x,\tau)h(\tau,t)d\tau. 
\end{equation}
To evaluate the characteristic function $\widetilde p(k,t)$ for $\alpha<2$ and $-\alpha<\theta<1$ 
we take into account only terms for small $|k|$. 
Expansion of $\widetilde p_0(k,\tau)$ produces the term $\tau^{c_\theta}$ and 
the integral from the function $\tau^{c_\theta}h(\tau,t)$, calculated by applying the expression 
for the Mellin transform from the Fox function, finally yields 
\begin{equation}
\label{sol}
p(x,t)={\cal N}L(|x|/A_F;\alpha,0),  
\end{equation}
where $A_F=c^{1/\alpha}t^{\beta/(\alpha+\theta)}\Gamma(c_\theta+1)^{1/\alpha}/\Gamma(\beta c_\theta+1)^{1/\alpha}$. 
Asymptotically, $p(x,t)\sim|x|^{-\alpha-1}$ and only moments of order $\delta<\alpha$ 
are finite. They directly follow from Eq.(\ref{sol}) and (\ref{mom0}) by applying 
the Mellin transform to Eq.(\ref{htt}): 
\begin{equation}
\label{mom}
\langle|x|^\delta\rangle(t)=\int |x|^\delta p_0(x,\tau)h(\tau,t)d\tau dx=A_M
t^{\delta\beta/(\alpha+\theta)}
\end{equation}
where $A_M=-\frac{2}{\pi\alpha\beta}A^{\delta/(\alpha+\theta)}\Gamma(-\delta/\alpha)\Gamma(1+\delta)
\Gamma(1+\delta/(\alpha+\theta))\sin(\pi\delta/2)/\Gamma(1+\delta\beta/(\alpha+\theta))$. 
The above time-dependence is also valid for $\theta\ge1$ and applying Eq.(\ref{var2}) 
yields $\langle x^2\rangle(t)\sim t^{2\beta/(2+\theta)}$ for the case $\alpha=2$. 

We will demonstrate that $p(x,t)$, Eq.(\ref{inte}), satisfies, in the limit of small $k$, 
the following integral equation which was phenomenologically introduced in Ref.\cite{sro07}, 
\begin{equation}
\label{gle}
\frac{\partial p(x,t)}{\partial t}=\int_0^t K(t-t')\frac{\partial^\alpha}{\partial|x|^\alpha}(|x|^{-\theta} p(x,t'))dt', 
\end{equation}
where the kernel $K$ is to be determined. For that purpose, we insert Eq.(\ref{inte}) into (\ref{gle}) and take 
the Fourier-Laplace transform neglecting the terms higher than $|k|^\alpha$, 
\begin{equation}
\label{glek}
{\cal L}[\partial \widetilde p(k,t)/\partial t]=-|k|^\alpha\widetilde K(u){\cal L}[\widetilde p_\theta(k,t)], 
\end{equation}
where the subscript $\theta$ means a multiplication by $|x|^{-\theta}$ and we inserted the Riesz fractional 
derivative ${\cal F}[\partial^\alpha/\partial|x|^\alpha]=-|k|^\alpha$. 
To evaluate the rhs of (\ref{glek}) in the diffusion limit, we need to take into account only the term $k^0$, 
i.e. to approximate ${\cal F}[|x|^{-\theta}p_0(x,\tau)]\approx c\tau^{-\theta/(\alpha+\theta)}$ \cite{sro14}. 
Then 
\begin{equation}
\label{pth}
\widetilde p_\theta(k,t)=c\int_0^\infty\tau^{-\theta/(\alpha+\theta)}h(\tau,t)d\tau
\end{equation}
and, using the expression for the Laplace transform with respect to $t$, 
$\widetilde h(\tau,u)=u^{\beta-1}\exp(-\tau u^\beta)$ \cite{wer}, we finally obtain 
\begin{equation}
\label{rhs}
{\cal L}[\widetilde p_\theta(k,t)]=c\Gamma\left(1-\frac{\theta}{\alpha+\theta}\right)u^{\theta\beta/(\alpha+\theta)-1}.
\end{equation}
The Fourier transformed lhs of (\ref{gle}) follows from the approximation 
$\widetilde p_0(k,\tau)=\exp(-c\tau^{c_\theta}|k|^\alpha)\approx 1-c\tau^{c_\theta}|k|^\alpha$:
\begin{equation}
\label{lhs}
\partial \widetilde p(k,t)/\partial t=-c|k|^\alpha\int_0^\infty\tau^{c_\theta}
\frac{\partial}{\partial t}h(\tau,t)d\tau.
\end{equation}
After taking the Laplace transform, we obtain the finite expression, 
\begin{equation}
\label{lhsf}
{\cal L}[\partial \widetilde p(k,t)/\partial t]=-c|k|^\alpha\Gamma(1+c_\theta)
u^{-\beta c_\theta}, 
\end{equation}
where we took into account that the integral vanishes at $t=0$ since it is proportional to 
$t^{\beta(c_\theta+1)}$. Inserting the expressions (\ref{rhs}) and (\ref{lhsf}) into Eq.(\ref{glek}) 
shows that the equation is satisfied and the kernel has the form 
\begin{equation}
\label{ker}
\widetilde K(u)=\frac{\Gamma\left(1+\frac{\alpha}{\alpha+\theta}\right)}
{\Gamma\left(1-\frac{\theta}{\alpha+\theta}\right)}u^{1-\beta}. 
\end{equation}
\begin{center}
\begin{figure}
\includegraphics[width=80mm]{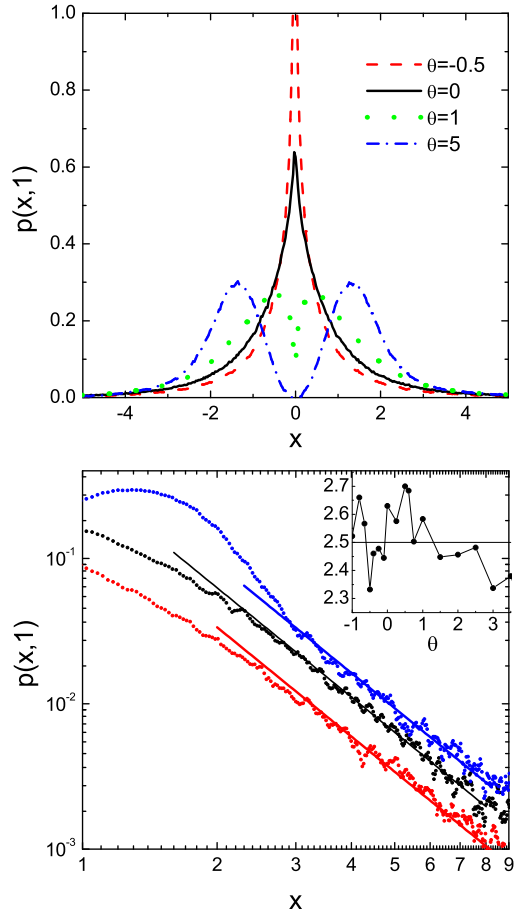}
\caption{(Colour online) Distributions obtained by numerical integration of Eq.(\ref{la}) 
with $g(x)=|x|^\theta$ for $\alpha=1.5$ 
and $\beta=0.5$. Points in the lower part correspond to $\theta=-0.8$, 0 and 5 (from bottom to top); 
the straight lines correspond to the slope -2.5. 
Inset: slopes fitted in the range $x\in(5,8)$. The ensemble involves $10^6$ events for each curve 
and the time step in the numerical integration was $10^{-4}$. }
\end{figure}
\end{center} 

It is convenient to express the integral equation (\ref{gle}) in terms of the Riemann-Liouville 
fractional operator, 
\begin{equation}
\label{rlo}
_0D_t^{1-\beta}f(t)=\frac{1}{\Gamma(\beta)}\frac{d}{dt}\int_0^t dt'\frac{f(t')}{(t-t')^{1-\beta}}.  
\end{equation}
For that purpose, we divide Eq.(\ref{glek}) by $u$, invert the transform using the formula 
${\cal L}[_0D_t^{-\beta}f(t)]=u^{-\beta}f(u)$ and differentiate over $t$. 
The final form of Eq.(\ref{gle}) involves two fractional operators, 
\begin{equation}
\label{glef}
\frac{\partial p(x,t)}{\partial t}=\frac{\Gamma(1+\frac{\alpha}{\alpha+\theta})}
{\Gamma(1-\frac{\theta}{\alpha+\theta})}{_0}D_t^{1-\beta}
\frac{\partial^\alpha}{\partial|x|^\alpha}(|x|^{-\theta} p(x,t)).
\end{equation}
Therefore, Eq.(\ref{las}) can be interpreted as a non-Markovian Langevin representation 
of the generalised Fokker-Planck equation (\ref{glef}). 
The case $\theta=0$ is well-known and then Eq.(\ref{glef}) follows from a simple scaling \cite{mag}. 
A version of Eq.(\ref{glef}), generalised to include a variable stability index, was derived
from CTRW in \cite{sti}. 

Eq.(\ref{la}) is manageable numerically by applying a modification of the standard method \cite{wer,kle} 
which allows us to determine $x(t)$ without performing an explicit time inversion $\tau(t)$. 
Fig.1 presents examples of the density distributions for some values of $\theta$ for both small and large $x$. 
The function $g(x)$ vanishes at 
$x=0$ for $\theta>0$ -- the dependence $t(\tau)$ is then very weak -- which results, in the physical time, 
in a repulsion of the particle from the centre. On the other hand, we observe a strong trapping 
in the origin for negative $\theta$. The tails fall like a power-law and the slope 
agrees with the value $-\alpha-1$, predicted by Eq.(\ref{sol}); deviations, shown in the inset, 
do not exceed the accuracy of the method. 
\begin{center}
\begin{figure}
\includegraphics[width=95mm]{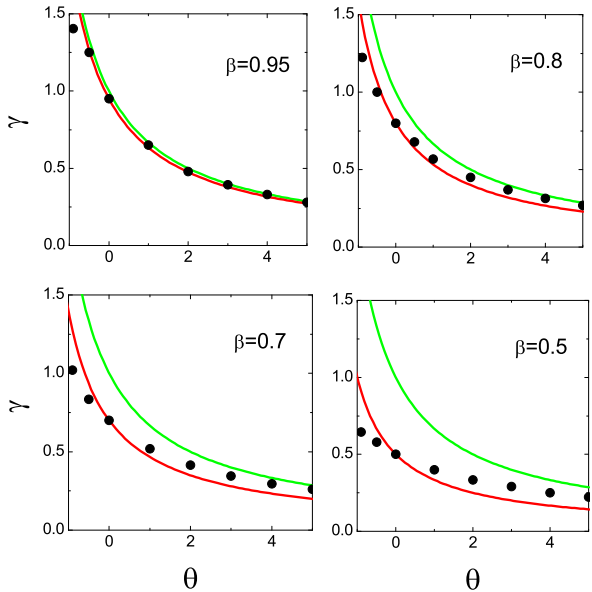}
\caption{(Colour online) Asymptotic time-dependence of the variance as a function of $\theta$ 
for $\alpha=2$ and some values of $\beta$, numerically calculated from Eq.(\ref{la}) (points). 
The upper green and the lower red lines mark the dependences $t^{2/(2+\theta)}$ and 
$t^{2\beta/(2+\theta)}$, respectively.}
\end{figure}
\end{center} 

Speed of the transport can be quantified by the moments and, for 
$\alpha=2$, the time-dependence of the variance determines whether the diffusion 
is normal or anomalous. In the latter case, according to Eq.(\ref{mom}), the parameter 
$\bar \gamma=2\beta/(2+\theta)$ is decisive and it satisfies the condition $0<\bar \gamma<2$. Therefore, all kinds of diffusion 
are predicted: a normal diffusion ($\bar \gamma=1$) and an anomalous one: sub- ($\bar \gamma<1$) or superdiffusion 
($\bar \gamma>1$). The above conclusion is qualitatively similar to the properties of 
the Markovian CTRW, Eq.(\ref{var2}); in this case we observe the same kinds of diffusion for 
$\theta=0$, $\theta>0$ and $-1<\theta<0$, respectively. 
The non-Markovian and time-homogeneous CTRW, in turn, predicts the subdiffusion for all $\beta<1$. 
Numerical integration of Eq.(\ref{la}), performed for $\alpha=2$, disclose a power-law rise of the variance, 
$\sim t^\gamma$, for large $t$; the index is presented in Fig.2 for a few values of $\beta$. 
Comparison of all those results points at the double origin of the long waiting times. 
For large $\theta$, the long waiting times at large distances are 
present already for the process $x(\tau)$ since the jumping rate for the underlying Markovian CTRW, 
$\sim|x|^{-\theta}$, becomes small. Consequently, the random component is from that point of view superfluous 
and the numerical $\gamma(\theta)$ actually converges to $2/(2+\theta)$ for $\theta\to\infty$. 
On the other hand, long, $\theta-$independent tails of $\xi$ determine the dynamics 
if $\theta$ is negative which results in a flat dependence $\gamma(\theta)$ for the numerical result. 
This is smaller than the value predicted by Eq.(\ref{mom}) and the enhanced diffusion turns into 
subdiffusion if $\beta$ becomes sufficiently small.


In conclusion, we proposed a formalism in which the subordinator became position-dependent 
by introducing a variable intensity $g(x)$. It is able to take into account spatial properties 
of the environment by a modification of the standard subordination procedure. 
Need for such a model is obvious when we consider complex, nonhomogeneous systems with traps 
and impurities. We numerically solved the Langevin equation, driven by the general L\'evy noise, 
for the subordinated process $x(\tau)$ assuming $g(x)$ in the algebraic form with the index $\theta$. 
The density distributions reflect the form of $g(x)$: a large $g$ correspond to a large density. 
Evaluation of the moments shows that, in the Gaussian case, all kinds of diffusion emerge but 
the enhanced diffusion vanishes for small $\beta$. 
The exact problem was approximated by representing the medium structure by 
the multiplicative Langevin equation in the operational time and a subsequent subordination of that 
process to the random time. The diffusion properties were derived; comparison with the numerical results 
demonstrates that the subordinated process determines the dynamics for very large $\theta$, whereas 
effects related to long waiting times dominate for negative $\theta$. The subordinating, 
multiplicative process $x(t)$ is governed by the double-fractional Fokker-Planck equation.


\begin{thebibliography}{99}

\bibitem{bou}
J.-P. Bouchaud and A. Georges, Phys. Rep. {\bf 195}, 127 (1990). 

\bibitem{met}
R. Metzler and J. Klafter, Phys. Rep. {\bf 339}, 1 (2000). 

\bibitem{fog}
H. C. Fogedby, Phys. Rev. E {\bf 50}, 1657 (1994). 

\bibitem{klag}
R. Klages, G. Radons, and I. M. Sokolov (Eds.), {\it Anomalous Transport: Foundations and Applications} 
(Wiley-VCH Verlag GmbH \& Co. KGaA, Weinheim, 2008). 

\bibitem{tall}
K. T. Tallakstad, R. Toussaint, S. Santucci, and K. J. M\aa l\o y, Phys. Rev. Lett. {\bf 110}, 145501 (2013). 

\bibitem{sch}
D. Schertzer, M. Larchev\^{e}que, J. Duan, V. V. Yanovsky, and S. Lovejoy, 
J. Math. Phys. {\bf 42}, 200 (2001).  

\bibitem{che} 
A. V. Chechkin, R. Gorenflo, and I. M. Sokolov, J. Phys. A: Math. Gen. {\bf 38}, L679 (2005).

\bibitem{sti}
B. A. Stickler and E. Schachinger, Phys. Rev. E {\bf 83}, 011122 (2011); 
{\it ibid} {\bf 84}, 021116 (2011). 

\bibitem{ben} 
E. Ben-Naim and P.L. Krapivsky, Phys. Rev. Lett. {\bf 102}, 190602 (2009). 

\bibitem{sro06}
T. Srokowski and A. Kami\'nska, Phys. Rev. E {\bf 74}, 021103 (2006). 

\bibitem{osh}
B. O'Shaughnessy and I. Procaccia, Phys. Rev. Lett. {\bf 54}, 455 (1985); 
R. Metzler, W. G. Gl\"ockle, and T. F. Nonnenmacher, Physica A {\bf 211}, 13 (1994). 

\bibitem{ved}
H. Fujisaka, S. Grossmann, and S. Thomae, Z. Naturforsch. Teil A {\bf 40}, 867 (1985); 
A. A. Vedenov, Rev. Plasma Phys. {\bf 3}, 229 (1967). 

\bibitem{sro14}
T. Srokowski, arXiv: cond-mat.stat-mech/1401.7175. 

\bibitem{pir}
A. Piryatinska, A. I. Saichev, and W. A. Woyczynski, Physica A {\bf 349}, 375 (2005).

\bibitem{non} 
R. Metzler and T. F. Nonnenmacher, Chem. Phys. {\bf 284}, 67 (2002). 

\bibitem{sro07}
T. Srokowski, Phys. Rev. E {\bf 75}, 051105 (2007).

\bibitem{wer}
M. Magdziarz, A. Weron, and K. Weron, Phys. Rev. E {\bf 75}, 016708 (2007).

\bibitem{mag}
M. Magdziarz and A. Weron, Phys. Rev. E {\bf 75}, 056702 (2007).

\bibitem{kle}
D. Kleinhans and R. Friedrich, Phys. Rev. E {\bf 76}, 061102 (2007). 

\end{thebibliography}
\end{document}